\documentclass[11pt,a4paper]{article}
\pdfoutput=1
\usepackage{jheppub}

\usepackage{url}
\usepackage{bm}

\allowdisplaybreaks[2]

\def\mint{\int_{-\infty}^\infty\!\cdots\!\int_{-\infty}^\infty}
\def\l{\ell}

\newcommand{\be}{\begin{equation}}
\newcommand{\ee}{\end{equation}}
\newcommand{\ba}{\begin{aligned}}
\newcommand{\ea}{\end{aligned}}

\def\({\left(}
\def\){\right)}

\DeclareMathOperator{\real}{Re}
\DeclareMathOperator{\im}{Im}

\DeclareMathOperator{\Hl}{H\ell}

\preprint{RUP-20-19}

\title{Quasinormal modes of Kerr-de Sitter black holes via the Heun function}

\author{Yasuyuki Hatsuda}

\affiliation{Department of Physics, Rikkyo University, Toshima, Tokyo 171-8501, Japan}

\emailAdd{yhatsuda@rikkyo.ac.jp}

\abstract{
This note addresses quasinormal mode frequencies of four-dimensional asymptotically de Sitter rotating black holes.
The main motivation is that \textit{Mathematica 12.1} has implemented a new family of special functions: \textit{Heun functions}. Using the fact that Teukolsky's equations for Kerr-de Sitter black holes are mapped to Heun's equations, we are able to compute their quasinormal mode frequencies by the Heun function.
In this approach, \textit{Mathematica} normally evaluates these frequencies to arbitrary numerical precision in a few seconds. 
We further discuss an application to asymptotically flat rotating black holes.
}

\begin{document}

\maketitle

\renewcommand{\thefootnote}{\arabic{footnote}}
\setcounter{footnote}{0}
\setcounter{section}{0}

\section{Motivations}
Recently, \textit{Mathematica} has implemented a new class of special functions, known as \textit{Heun functions} \cite{zotero-803}.
The Heun function is a solution to Heun's differential equation:
\begin{equation}
\begin{aligned}
y''(z)+\( \frac{\gamma}{z}+\frac{\delta}{z-1}+\frac{\epsilon}{z-a} \) y'(z)+\frac{ \alpha \beta z-q}{z(z-1)(z-a)}y(z)=0,
\end{aligned}
\label{eq:Heun}
\end{equation}
where $\gamma+\delta+\epsilon=\alpha+\beta+1$. 
Heun's equation is a second order Fuchsian differential equation with four singular points \cite{ronveaux1995, kristensson2010, zotero-804}. The Heun function is a generalization of the Gauss hypergeometric function, but has some significant differences. 
For instance, it has an additional parameter $q$, called an accessory parameter, which does not affect the charcteristic exponents at any singular point.
Since the Heun function is much less familiar than the hypergeometric function, \textit{Mathematica} did not have it for a long time.%
\footnote{It seems that \textit{Maple} had already implemented it \cite{zotero-819, fiziev2006}. Unfortunately the author has never used \textit{Maple}. We were also informed that a numerical code in \textit{GNU Octave} is publicly available \cite{motygin2015}.}

Motivated by this great update,
we apply this special function to black hole physics. For other possible applications, see \cite{birkandan2017, hortacsu2018a} for instance. The idea is very simple. It is known that both angular and radial Teukolsky equations \cite{teukolsky1972} for Kerr (anti-)de Sitter black holes are mapped to Heun's differential equations \cite{suzuki1998}. Their solutions are expressed by the Heun function. 
Using them, we compute quasinormal mode (QNM) frequencies as well as generalized spin-weighted spheroidal eigenvalues.

The QNM problem for the Kerr-(A)dS$_4$ black holes has a great advantage.
Since all the singular points in Heun's equation are regular, these are all equivalent.
If one constructs local solutions at a certain singular point, one can use them in the construction of local solutions at the other singular points by symmetry.
In fact, it is well-known that all the local solutions in Heun's equation can be expressed in terms of the same (local) Heun function,%
\footnote{There are a huge number of symmetries among the local solutions \cite{maier2007}. It is unclear for us what such symmetries imply for black hole physiscs.}
as is so in the hypergeometric differential equation.
This nice property, however, is lost if one considers asymptotically flat geometries. In these cases, two regular singular points get confluent, and result in an irregular singular point. Connection problems between a regular singular point and an irregular singular point are more difficult in general.
One needs to modify the argument here for such problems \cite{fiziev2006, fiziev2010}. 
We will be back to this issue in the concluding section.

We explicitly construct local solutions to the Teukolsky equation for the Kerr-dS$_4$ black holes, and use them to determine the QNM frequencies.
We use the standard Wronskian method to do so. It is easy to put all the computations on \textit{Mathematica}, and our simple program
provides the QNM frequencies very rapidly (in a few seconds) and precisely (up to any digits). 
We do them for the Schwarzschild-dS and Kerr-dS cases.
All the obtained results are consistent with available data in the literature.

\paragraph{\textit{Implementation.}}
The reader finds a sample \textit{Mathematica} notebook in arXiv.
This program works only in \textit{Mathematica} \textit{12.1} (or later).
All the numerical data in this note are simply generated by this program.

\section{From Teukolsky to Heun}
Since our purpose is merely to show an efficiency of the Heun function in \textit{Mathematica},
we do not discuss perturbations of the Kerr-(A)dS black holes. 
We start our argument with the Teukolsky equation for such black holes \cite{khanal1983, chambers1994, suzuki1998}.
The Kerr-(A)dS$_4$ black holes have three independent parameters: mass $M$, angular momentum per unit mass $a$ and the cosmological constant $\Lambda$.
For later convenience, we define a new parameter by
\begin{equation}
\begin{aligned}
\alpha=\frac{\Lambda a^2}{3}.
\end{aligned}
\end{equation}
To fix our convention, let us write the metric of the Kerr-(A)dS$_4$ black holes in the Boyer-Lindquist coordinates:
\begin{equation}
\begin{aligned}
ds^2=-\frac{\Delta_r}{(1+\alpha)^2\rho^2}(dt-a\sin^2\theta d\phi)^2+\frac{\Delta_\theta \sin^2 \theta}{(1+\alpha)^2 \rho^2}
[a dt-(r^2+a^2)d\phi]^2+\frac{\rho^2}{\Delta_r}dr^2+\frac{\rho^2}{\Delta_\theta} d\theta^2,
\end{aligned}
\end{equation}
where $\rho^2=r^2+a^2\cos^2\theta$ and
\begin{equation}
\begin{aligned}
\Delta_r=(r^2+a^2)\(1-\frac{\Lambda}{3}r^2\)-2M r,\qquad
\Delta_\theta=1+\alpha \cos^2\theta.
\end{aligned}
\end{equation}

\subsection{Angular part}
We basically follow the convention in \cite{suzuki1998}. 
The Teukolsky equation is separated into the angular and the radial equations.
The angular Teukolsky equation is given by eq. (3.1) in \cite{suzuki1998}. We rewrite it as a more compact form:
\begin{equation}
\begin{aligned}
&\biggl[ \frac{d}{dx} (1+\alpha x^2)(1-x^2) \frac{d}{dx}+ \lambda-s(1-\alpha)-2\alpha x^2\\
&\quad+\frac{(1+\alpha)^2}{1+\alpha x^2}\biggl( c^2 x^2-2csx-c^2+2cm 
+\frac{4\alpha}{1+\alpha}smx-\frac{(m+sx)^2}{1-x^2} \biggr)
 \biggr] S(x)=0,
\end{aligned}
\label{eq:angular}
\end{equation}
where $x=\cos \theta$, $s$ is a spin-weight of a perturbing field, $m$ is an azimuthal number and $c=a\omega$ with frequency $\omega$.
$\lambda$ is a separation constant, which is determined by a requirement of the regularity for $S(x)$ both at $x=\pm 1$.
We will show that this two-point boundary value problem is solved by the Heun function.

In the flat limit $\Lambda \to 0$ (i.e., $\alpha \to 0$), the angular equation \eqref{eq:angular} reduces to the so-called spin-weighted spheroidal equation \cite{teukolsky1972}.
The relation to the original separation constant $A$ in \cite{teukolsky1972} is $\lambda|_{\alpha=0}=A+2s-2cm+c^2$.
Even in this case, the eigenvalue is not known analytically.\footnote{Very recently, Aminov, Grassi and the author conjectured its analytic expression in terms of a special function in an $\mathcal{N}=2$ SU(2) supersymmetric QCD \cite{aminov2020}.} If taking a further limit $a \to 0$, then $\lambda \to \l(\l+1)-s(s-1)$ with multipole number $\l$.

If we take the spinless limit $a \to 0$ first with $\omega$ and $\Lambda$ fixed (i.e., $c \to 0$ and $\alpha \to 0$), then the reduced equation does not contain $\Lambda$. Therefore we conclude $\lambda|_{a=0}=\l(\l+1)-s(s-1)$. 

At first glance, the differential equation \eqref{eq:angular} has five regular singular points at $x=\pm1, \pm i/\sqrt{\alpha}, \infty$.
However the infinity point turns out to be a removable singularity. Its characteristic exponents are $-1$ and $0$, and moreover the corresponding two Frobenius solutions have no logarithmic singularities. 
As a consequence, it is removed by a proper transformation of $S(x)$.
The angular differential equation finally leads to Heun's differential equation.
This surprising fact was first shown in \cite{suzuki1998}. We follow their argument.
To see the relation, we first match locations of the singular points.
This is done by a M\"obius transformation:\footnote{There are $4!$ possible M\"obius transformations. We take one of them so that our interested boundary condition is imposed at $z=0$ and $z=1$.}
\begin{equation}
\begin{aligned}
z=\frac{(1-i/\sqrt{\alpha})(x+1)}{2(x-i/\sqrt{\alpha})}.
\end{aligned}
\end{equation}
In this transformation, $x=-1, 1, -i/\sqrt{\alpha}, i/\sqrt{\alpha}, \infty$ are mapped into $z=0, 1, z_a, \infty, z_{\infty}$, respectively,
where
\begin{equation}
\begin{aligned}
z_a:= -\frac{(1-i/\sqrt{\alpha})^2}{4i/\sqrt{\alpha}},\qquad
z_{\infty}:=\frac{1-i/\sqrt{\alpha}}{2}.
\end{aligned}
\end{equation}
We next perform the following transformation of $S(x)$:
\begin{equation}
\begin{aligned}
S(x)=z^{A_1}(z-1)^{A_2}(z-z_a)^{A_3} (z-z_{\infty}) y_a(z),
\end{aligned}
\end{equation}
where 
\begin{equation}
\begin{aligned}
A_1=\frac{m-s}{2},\qquad A_2=-\frac{m+s}{2},\qquad
A_3=\frac{i}{2}\( \frac{1+\alpha}{\sqrt{\alpha}}c-m\sqrt{\alpha} -is \).
\end{aligned}
\end{equation}
The new function $y_a(z)$ now satisfies 
\begin{equation}
\begin{aligned}
y_a''(z)+\( \frac{2A_1+1}{z}+\frac{2A_2+1}{z-1}+\frac{2A_3+1}{z-z_a} \)y_a'(z)+\frac{\rho_+\rho_- z+u}{z(z-1)(z-z_a)} y_a(z)=0,
\end{aligned}
\label{eq:Heun-a}
\end{equation}
where
\begin{equation}
\begin{aligned}
\rho_+&=1,\qquad \rho_-=1-s-im\sqrt{\alpha}+ic\( \sqrt{\alpha}+\frac{1}{\sqrt{\alpha}} \), \\
u&=-\left[ \frac{i\lambda}{4\sqrt{\alpha}}+\frac{1}{2}+A_1+\( m+\frac{1}{2} \) (A_3-A_3^*) \right].
\end{aligned}
\end{equation}
Here a `conjugate' $A_3^*$ is obtained by replacing $i$ in $A_3$ by $-i$.
Now the point $z=z_{\infty}$ is not a singular point as expected.
Note that there is a relation: $(2A_1+1)+(2A_2+1)+(2A_3+1)=\rho_++\rho_-+1$.
Therefore the differential equation \eqref{eq:Heun-a} is indeed Heun's differential equation whose solutions are discussed later.

\subsection{Radial part}
The radial Teukolsky equation is more complicated.
It is given by\footnote{There seems a small misprint in eq. (4.3) in \cite{khanal1983}. We have fixed it, following \cite{suzuki1998}. It is consistent with \cite{chambers1994, giammatteo2005} after matching the conventions.}
\begin{equation}
\begin{aligned}
\biggl[ \Delta_r^{-s} \frac{d}{dr} \Delta_r^{s+1} \frac{d}{dr} +\frac{(1+\alpha)^2 K^2-is(1+\alpha)K \Delta_r'}{\Delta_r}
+4is(1+\alpha)\omega r\\
-\frac{2\alpha}{a^2}(s+1)(2s+1)r^2+2s(1-\alpha)-\lambda \biggr] R(r)=0,
\end{aligned}
\label{eq:radial}
\end{equation}
where
\begin{equation}
\begin{aligned}
K(r)=\omega(r^2+a^2)-am.
\end{aligned}
\end{equation}
For the de Sitter black holes, $\Delta_r=0$ has four real roots. We label these roots $r_\pm$ and $r_\pm'$ so that $r_-' < r_-  < r_+ < r_+'$.
As in the angular case, the radial Tuekolsky equation has five regular singular points $r=r_{\pm}, r_{\pm}', \infty$,
but the singularity at infinity is apparent and removable.
To go to Heun's world, we need a non-trivial transformation.
We do a M\"obius transformation:
\begin{equation}
\begin{aligned}
z=\frac{(r_+'-r_-)(r-r_+)}{(r_+'-r_+)(r-r_-)}.
\end{aligned}
\end{equation}
After it, the five points are mapped as
\begin{equation}
\begin{aligned}
(r_+,r_+', r_-, r_-', \infty) \to (0,1, \infty, z_r, z_\infty),
\end{aligned}
\end{equation}
where
\begin{equation}
\begin{aligned}
z_r:=\frac{(r_+'-r_-)(r_-'-r_+)}{(r_+'-r_+)(r_-'-r_-)},\qquad z_\infty:=\frac{r_+'-r_-}{r_+'-r_+}.
\end{aligned}
\end{equation}
Note that $z_r>1$ for the Kerr-dS black holes.
Following \cite{suzuki1998}, we further transform
\begin{equation}
\begin{aligned}
R(r)=z^{B_1}(z-1)^{B_2}(z-z_r)^{B_3}(z-z_\infty)^{2s+1} y_r(z),
\end{aligned}
\end{equation}
where
\begin{equation}
\begin{aligned}
B_1=\frac{i(1+\alpha)K(r_+)}{\Delta_r'(r_+)},\quad
B_2=\frac{i(1+\alpha)K(r_+')}{\Delta_r'(r_+')},\quad
B_3=\frac{i(1+\alpha)K(r_-')}{\Delta_r'(r_-')}.
\end{aligned}
\label{eq:B}
\end{equation}
Then, we finally obtain the following equation:
\begin{equation}
\begin{aligned}
y_r''(z)+\( \frac{2B_1+s+1}{z}+\frac{2B_2+s+1}{z-1}+\frac{2B_3+s+1}{z-z_r} \)y_r'(z)
+\frac{\sigma_+\sigma_- z+v}{z(z-1)(z-z_r)} y_r(z)=0,
\end{aligned}
\label{eq:Heun-r}
\end{equation}
where
\begin{equation}
\begin{aligned}
\sigma_+=2s+1,\qquad
\sigma_-=s+1-\frac{2i(1+\alpha)K(r_-)}{\Delta_r'(r_-)},
\end{aligned}
\label{eq:sigma}
\end{equation}
and
\begin{equation}
\begin{aligned}
v=
\frac{(1+s)(1+2s)r_-'}{r_--r_-'}+\frac{3\lambda-2s(3-a^2 \Lambda)+\Lambda (1+s)(1+2s) r_+(r_++r_+')}{\Lambda(r_--r_-')(r_+-r_+')} \\
-\frac{2i(1+2s)(3+a^2 \Lambda)(r_+ r_- \omega+a^2 \omega-a m)}{\Lambda(r_--r_-')(r_--r_+)(r_+-r_+')}.
\end{aligned}
\label{eq:v}
\end{equation}
The expression of $v$ looks much simpler than that in \cite{suzuki1998} for $Q=0$. 
Note that to derive \eqref{eq:Heun-r}, we do not need explicit forms of $r_\pm, r_\pm'$ in terms of $M$, $a$ and $\Lambda$.
We have used a trivial relation $r_++r_+'+r_-+r_-'=0$ and the following nice identity:
\begin{equation}
\begin{aligned}
\frac{K(r_+)}{\Delta_r'(r_+)}+\frac{K(r_+')}{\Delta_r'(r_+')}+\frac{K(r_-)}{\Delta_r'(r_-)}+\frac{K(r_-')}{\Delta_r'(r_-')}=0.
\end{aligned}
\end{equation}
Again the equation \eqref{eq:Heun-r} is Heun's equation.

\subsection{The Heun function}
In the previous subsections we showed that the Teukolsky equations \eqref{eq:angular} and \eqref{eq:radial} are transformed into Heun's differential equations \eqref{eq:Heun-a} and \eqref{eq:Heun-r}, respectively.
Here we look at their local solutions at singular points.

Let us consider Heun's equation \eqref{eq:Heun}.
There are a few conflicts of letters in the previous subsections, but they will cause no confusions.
As already mentioned, Heun's equation has four regular singular points at $z=0,1,a,\infty$.
For each point, we can construct Frobenius solutions.
They are convergent inside a circle, whose radius is generically determined by the distance from the nearest singularity.
Since the exponents at $z=0$ are $0$ and $1-\gamma$,
let $\Hl(a,q;\alpha,\beta,\gamma,\delta;z)$ be the regular solution at $z=0$.
We normalize it as
\begin{equation}
\begin{aligned}
\Hl(a,q;\alpha,\beta,\gamma,\delta;0)=1.
\end{aligned}
\label{eq:norm}
\end{equation}
We refer to this local solution as \textit{the Heun function}.\footnote{A technical remark: The term ``Heun function'' has different meanings in the literature. In \cite{ronveaux1995} this term is used for a solution that is analytic at two singular points. That is, it is a solution satisfying a two-point boundary condition. This condition requires a continued fraction equation for the accessory parameter \cite{ronveaux1995}, which is compared with Leaver's approach \cite{leaver1985} in black hole perturbations.  
This terminology seems standard. In \textit{Mathematica}, the same term seems to mean the regular solution at the origin, which is called merely a local solution in \cite{ronveaux1995}. In this note, we will use it in the latter sense for simplicity though it is not conventional. 
} 
It turns out that the two solutions at $z=0$ are given by \cite{zotero-804},
\begin{equation}
\begin{aligned}
y_{01}(z)&=\Hl(a,q;\alpha,\beta,\gamma,\delta;z),\\
y_{02}(z)&=z^{1-\gamma} \Hl(a, (a\delta+\epsilon)(1-\gamma)+q;\alpha+1-\gamma,\beta+1-\gamma,2-\gamma,\delta;z).
\end{aligned}
\label{eq:sol-z0}
\end{equation}
Unless $\gamma$ is an integer, these two are independent.
Since in our application, the singular point $z=a$ is always located outside the unit cirecle: $|a|>1$, the radius of convergence of these local solutions is unity in general.

Similarly, two solutions at $z=1$ are given by
\begin{align}
&y_{11}(z)=\Hl(1-a,\alpha\beta-q;\alpha,\beta,\delta,\gamma;1-z), \notag \\
&y_{12}(z)=(1-z)^{1-\delta} \label{eq:sol-z1}\\
&\qquad\quad \times \Hl(1-a, ((1-a)\gamma+\epsilon)(1-\delta)+\alpha\beta-q;\alpha+1-\delta,\beta+1-\delta,2-\delta,\gamma;1-z). \notag
\end{align}
The other solutions near $z=a$ and $z=\infty$ can be also expressed in terms of $\Hl$. They are not needed in our analysis.

In two-point boundary value problems between $z=0$ and $z=1$, it is useful to consider connection relations among the local solutions in different domains.
Let us write them as
\begin{equation}
\begin{aligned}
y_{01}(z)&=C_{11} y_{11}(z)+C_{12} y_{12}(z), \\
y_{02}(z)&=C_{21} y_{11}(z)+C_{22} y_{12}(z),
\end{aligned}
\label{eq:connection}
\end{equation}
where we have assumed that the solutions $y_{11}(z)$ and $y_{12}(z)$ are linearly independent.
In the angular eigenvalue problem, this assumption is often violated. Nevertheless the Wronskian method still works. 

To our knowledge, unlike the hypergeometric function, no analytically explicit representations of the connection coefficients $C_{ij}$ are known for the Heun solutions.
However, formally these are represented by Wronskians of local solutions, for instance, $C_{12} =W[y_{01}, y_{11}]/W[y_{12}, y_{11}]$ etc.
Therefore, if we know the local solutions, we can evaluate the connection coefficients.
This point has just become doable in the recent update of \textit{Mathematica}.

In \textit{Mathematica}, the Heun function \cite{zotero-803} is implemented as
\begin{equation}
\begin{aligned}
\texttt{HeunG[a,q,$\bm{\alpha}$,$\bm{\beta}$,$\bm{\gamma}$,$\bm{\delta}$,z]}\,.
\end{aligned}
\end{equation}
Recall that $\epsilon=1+\alpha+\beta-\gamma-\delta$ is not an independent parameter.
The function `\texttt{HeunG}' satisfies Heun's differential equation \eqref{eq:Heun} and has the same normalization \eqref{eq:norm}.
Therefore locally `\texttt{HeunG}' is exactly the same as $\Hl$.
Globally `\texttt{HeunG}' gives analytic continuation of $\Hl$. Though sometimes multi-valuedness causes a problem in numerical computations,
all our computations are done inside the convergence circle. There is no multi-value problem.

\section{From Heun to QNMs}
We are ready to apply the Heun function to the computations of the QNM frequencies of the Kerr-dS$_4$ black holes.

\subsection{Schwarzschild-de Sitter}
As a warm-up, we begin with non-rotating asymptotically dS$_4$ black holes.
In the non-rotating cases, the angular problem is trivialized, and we have
\begin{equation}
\begin{aligned}
\lambda|_{a=0}=\l(\l+1)-s(s-1).
\end{aligned}
\end{equation}
We can take the limit $a \to 0$ smoothly in \eqref{eq:B}, \eqref{eq:sigma} and \eqref{eq:v}.
Also, for $a=0$, the inner horizon is always at $r_-=0$.

Now two solutions near the event horizon $z=0$ ($r=r_+$) are given by \eqref{eq:sol-z0} with the identification
\begin{equation}
\begin{aligned}
a&=z_r,\qquad q=-v,\qquad \alpha=\sigma_+,\qquad \beta=\sigma_-,\\
\gamma&=2B_1+s+1,\quad \delta=2B_2+s+1,\quad \epsilon=2B_3+s+1.
\end{aligned}
\label{eq:Heun-para}
\end{equation}
Similarly, the solutions near the de Sitter horizon $z=1$ ($r=r_+'$) are given by \eqref{eq:sol-z1} with the same correspondence.

To compute the QNM frequencies, we have to see asymptotic behaviors.
Near $z=0$, the two solutions behave as $y_{01} \sim z^0$, $y_{02} \sim z^{1-\gamma}$. Therefore the original radial eigenfunctions show
\begin{equation}
\begin{aligned}
R_{01}(r):=z^{B_1}(z-1)^{B_2}(z-z_r)^{B_3}(z-z_\infty)^{2s+1} y_{01}(z) &\sim (r-r_+)^{i\omega r_+^2/\Delta_r'(r_+)} ,\\
R_{02}(r):=z^{B_1}(z-1)^{B_2}(z-z_r)^{B_3}(z-z_\infty)^{2s+1} y_{02}(z) &\sim (r-r_+)^{-s-i\omega r_+^2/\Delta_r'(r_+)}.
\end{aligned}
\end{equation}
We conclude that in our convention $R_{02}(r)$ satisfies the QNM boundary condition at the event horizon.
Similarly, near $z=1$, we have
\begin{equation}
\begin{aligned}
R_{11}(r):=z^{B_1}(z-1)^{B_2}(z-z_r)^{B_3}(z-z_\infty)^{2s+1} y_{11}(z) &\sim (r_+'-r)^{i\omega {r_+'}^2/\Delta_r'(r_+')} ,\\
R_{12}(r):=z^{B_1}(z-1)^{B_2}(z-z_r)^{B_3}(z-z_\infty)^{2s+1} y_{12}(z) &\sim (r_+'-r)^{-s-i\omega {r_+'}^2/\Delta_r'(r_+')}.
\end{aligned}
\end{equation}
Therefore $R_{11}(r)$ is a preferred solution.
The QNM boundary condition then requires $C_{22}=0$ in \eqref{eq:connection}.
This leads to the vanishing condition of the Wronskian:
\begin{equation}
\begin{aligned}
C_{22}(\omega)=\frac{W[y_{02}, y_{11}]}{W[y_{12}, y_{11}]}=0.
\end{aligned}
\end{equation}
In the evaluation of the Wronskian, we choose a value of $z$ so that it is located in both convergence circles at $z=0$ and $z=1$.
Typically $z=1/2$ is a good point for this purpose. The ratio of the Wronskian is independent of $z$.
Solving this equation, we get the QNM frequencies $\omega$.

For given $(M, \Lambda)$ and $(s,\l)$, we compute all the quantities in \eqref{eq:Heun-para}, and evaluate the Wronskian as a function of $\omega$ by using `\texttt{HeunG}.'
The result for $(s,\l)=(2,2)$ is shown in Table~\ref{tab:SdS}.
In this approach, the eigenvalues are obtained in a few seconds.
To find a solution to $C_{22}(\omega)=0$, we have to choose an initial value of `\texttt{FindRoot}' carefully.
Such an initial value is estimated by using semi-analytic results or by referring to a previous value if parameters are varied slightly. 
Other numerical codes such as \cite{jansen2017, hatsuda2020} are also useful for this purpose.
To make Table~\ref{tab:SdS}, we slightly varied the cosmological constant, and used a previous eigenvalue as a next initial value.

Strictly speaking, the results here are obtained by the non-rotating limit of the Teukolsky equation.
The result in \cite{zhidenko2003} is based on the Regge--Wheeler equation. Therefore our test also confirms the isospectrality of these distinct master equations.

We should also refer to scalar perturbations. The result in \cite{zhidenko2003} is for the minimally coupled massless scalar perturbation, while
the result in this note is for the conformally coupled massless scalar perturbation. The minimally coupled scalar perturbation cannot be mapped into Heun's equation, as was shown in \cite{aminov2020}. It is possible if and only if the massless scalar is conformally coupled to the Ricci scalar.

\begin{table}[tp]
\caption{The quasinormal mode frequencies of the Schwarzschild dS$_4$ black holes for $(s,\l)=(2,2)$. We show the results for the lowest and next lowest overtone modes. They are compared with tables in \cite{zhidenko2003}.}
\begin{center}
\begin{tabular}{ccc}
\hline
$M^2 \Lambda$ & $M \omega_0$ & $M \omega_1$ \\
\hline
$0.02$ & $0.33839142759 - 0.08175644548i$ & $0.31875866800 - 0.24919663091i$ \\
$0.04$ & $0.29889472383 - 0.07329667837i$ & $0.28584093744 - 0.22172414818i$ \\
$0.06$ & $0.25328922296 - 0.06304252803i$ & $0.24574199932 - 0.18979104298i$ \\
$0.08$ & $0.19748226379 - 0.04987732867i$ & $0.19411482671 - 0.14978664945i$ \\
$0.10$ & $0.117916433909 - 0.030210488607i$ & $0.11724321914 - 0.09064094839i$ \\
$0.11$ & $0.037269946073 - 0.009615651532 i$ & $0.037249337619 - 0.028846981555i$ \\
\hline
\end{tabular}
\end{center}
\label{tab:SdS}
\end{table}%

As a further benchmark, we compare with a high-precision method recently proposed in \cite{hatsuda2020}.
In this approach, numerical precisions get better for smaller overtone numbers and for larger multipole numbers.
We set $(s,\ell)=(2,5)$, and compute the lowest overtone frequency for $M^2 \Lambda=1/100, 1/10$.
The two distinct approaches showed at least 50-digit agreements.
A numerical limitation occurs in the approach of \cite{hatsuda2020}, but the method in this note does not seem to show such a limitation.

\subsection{Kerr-de Sitter}
The extension to the Kerr black holes is very simple.
The radial problem is almost the same as the Schwarzschild case.
We consider the angular problem.
Two solutions near $z=0$ ($x=-1$) and $z=1$ ($x=1$) are given by \eqref{eq:sol-z0} and \eqref{eq:sol-z1}, respectively.
We need the following identification:
\begin{equation}
\begin{aligned}
a&=z_a,\qquad q=-u, \qquad \alpha=\rho_+,\qquad \beta=\rho_- ,\\
\gamma&=2A_1+1,\qquad \delta=2A_2+1,\qquad \epsilon=2A_3+1.
\end{aligned}
\end{equation}
Let us see the boundary conditions. Near $x=-1$ ($z=0$), the original eigenfunctions behave as
\begin{equation}
\begin{aligned}
S_{01}(x)&:=z^{A_1}(z-1)^{A_2}(z-z_a)^{A_3}(z-z_{\infty}) y_{a01}(z) \sim (1+x)^{(m-s)/2} ,\\
S_{02}(x)&:=z^{A_1}(z-1)^{A_2}(z-z_a)^{A_3}(z-z_{\infty}) y_{a02}(z) \sim (1+x)^{-(m-s)/2}.
\end{aligned}
\end{equation}
where we use the subscript $a$ to distinguish the solutions from those for the radial problem.
Hence we should choose $S_{01}(x)$ for $m > s$ or $S_{02}(x)$ for $m < s$. For $m=s$, both are equivalent.
Similarly, the behavior near $x=1$ is given by
\begin{equation}
\begin{aligned}
S_{11}(x)&:=z^{A_1}(z-1)^{A_2}(z-z_a)^{A_3}(z-z_{\infty}) y_{a11}(z) \sim (1-x)^{-(m+s)/2} ,\\
S_{12}(x)&:=z^{A_1}(z-1)^{A_2}(z-z_a)^{A_3}(z-z_{\infty}) y_{a12}(z) \sim (1-x)^{(m+s)/2}.
\end{aligned}
\end{equation}
The regularity condition depends on the sign of $m+s$.
In summary, we have four branches to classify the regularity conditions of $S(x)$ depending on values of $s$ and $m$.
The same classification is found in the asymptotically flat case in \cite{fiziev2010}.

Let us consider the case $(s,\l)=(-2,2)$ for concreteness. Since $|m| \leq 2$, the regularity condition requires
$S_{01}(x)$ and $S_{11}(x)$ at both sides. 
Their linear dependence leads to $C_{12}=0$. However, in the angular problem, the two exponents $0$ and $1-\delta$ at $z=1$ are integers.
Therefore the linear independence of $y_{a11}(z)$ and $y_{a12}(z)$ is not guaranteed.
Recall that our interested boundary condition is just the regularity of $S(x)$ at $x=\pm 1$.
This condition is simply satisfied by the linear dependence of $S_{01}(x)$ and $S_{11}(x)$, which leads to
\begin{equation}
\begin{aligned}
W[y_{a01}, y_{a11}]=0.
\end{aligned}
\label{eq:angular-cond}
\end{equation}
Keep in mind that this condition may be modified, depending on the values of $s$ and $m$.
In the radial problem, we impose the same condition as in the previous subsection:
\begin{equation}
\begin{aligned}
\frac{W[y_{r02}, y_{r11}]}{W[y_{r12}, y_{r11}]}=0.
\end{aligned}
\label{eq:radial-cond}
\end{equation}
These two conditions determine $\omega$ and $\lambda$ simultaneously.
We show some numerical results in Table~\ref{tab:Kerr}.

\begin{table}[t]
\caption{The eigenvalues of the Kerr-dS$_4$ black hole for $(s,\l,m)=(-2,2,2)$. For comparison, we borrowed a setup in \cite{yoshida2010}.  The cosmological constant is fixed at $\Lambda=3$. The mass $M(\approx 0.1205)$ is fixed precisely so that $\Delta_r'(r_+)/(2r_+)|_{a=0}=4/10$. The results are compared with Table I in \cite{yoshida2010}.}
\begin{center}
\begin{tabular}{ccc}
\hline
$a$  &  ${}_{-2}\omega_{22}$  &  ${}_{-2}\lambda_{22}$  \\
\hline
$10^{-4}$  &  $2.418767521 - 0.594425984i$  & $-0.001612340016 + 0.000396276388i$ \\
$0.05$ &  $3.120126838 - 0.582550072i$  &  $-0.9978941169 + 0.1924823086 i$ \\
$0.1$  &  $4.545705311 - 0.512759887i$ &  $-2.860148411 + 0.335152309 i$ \\ 
$0.121$ & $6.785133281 - 0.221854155i$  &  $-5.186821705 + 0.173003125i$  \\
$10^{-4}$  &  $-2.416524801 - 0.594456639i$  &  $0.001611188212 + 0.000396312032i$ \\
$0.05$  &  $-1.940643745 - 0.602211690i$  &  $0.6904947105 + 0.2029675995i$  \\
$0.1$  &  $-1.579725415 - 0.612609462 i$  &  $1.2295440010 + 0.4190463078i$ \\
$0.121$  &  $-1.451660548 - 0.617989918i$  &  $1.430082521 + 0.515303128i$  \\
\hline
\end{tabular}
\end{center}
\label{tab:Kerr}
\end{table}%

\section{Remarks}
In this note, we reported the power of `\texttt{HeunG}' in \textit{Mathematica 12.1} to get quasinormal frequencies of Kerr-de Sitter black holes.
This way enables us to reach the QNM frequencies very accurately in a remarkably short time. We can also construct the explicit eigenfunctions. To search numerical solutions to the Wronskian conditions, we need proper initial values. So far, we do not have a systematic way to do so for the Kerr-dS black holes. It is desirable to find (semi-)analytic results for this purpose. Some partial results are found in \cite{suzuki1998, giammatteo2005, novaes2019} for example.

In the Kerr-AdS$_4$ black holes, the construction of the solutions is the same as the dS case.
In the computation of the QNMs, however, the boundary condition is more involved. 
The boundary condition for the radial Teukolsky equation of the Kerr-AdS$_4$ black holes was discussed in \cite{giammatteo2005}.
It would be nice to apply our method to the computation of the QNM eigenvalues for the Kerr-AdS case.

The flat case is more delicate in a sense. In the radial problem we have to consider a two-point boundary problem between a regular singular point and an irregular singular point. The confluent Heun function `\texttt{HeunC}' is based on the regular solution at $z=0$, a regular singular point. We cannot use it for the construction of local solutions at $z=\infty$, the irregular singular point. We need modifications to compute the QNMs for the flat case.

\begin{figure}[tb]
\begin{center}
  \begin{minipage}[b]{0.45\linewidth}
    \centering
    \includegraphics[width=0.95\linewidth]{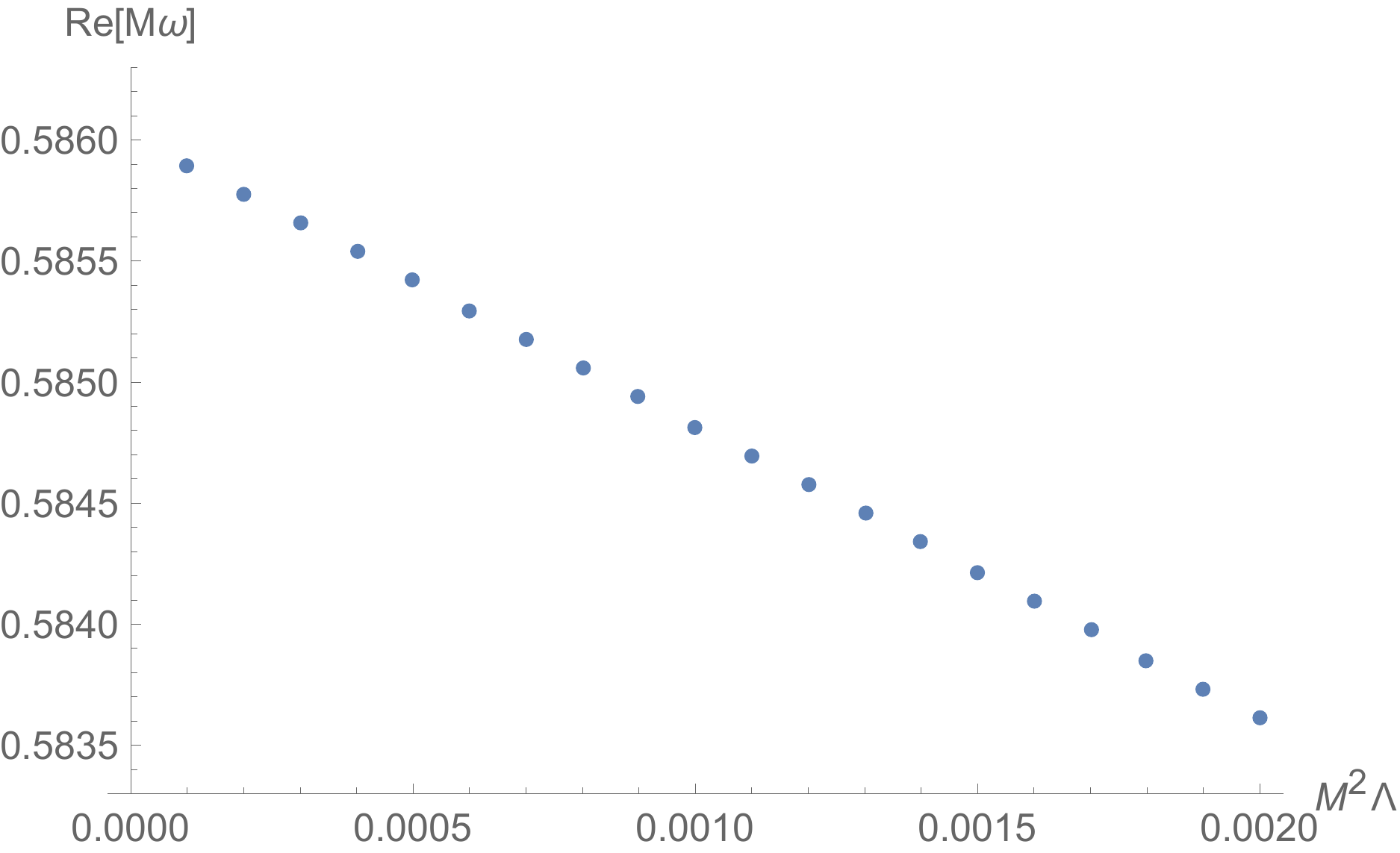}
  \end{minipage} \hspace{1cm}
  \begin{minipage}[b]{0.45\linewidth}
    \centering
    \includegraphics[width=0.95\linewidth]{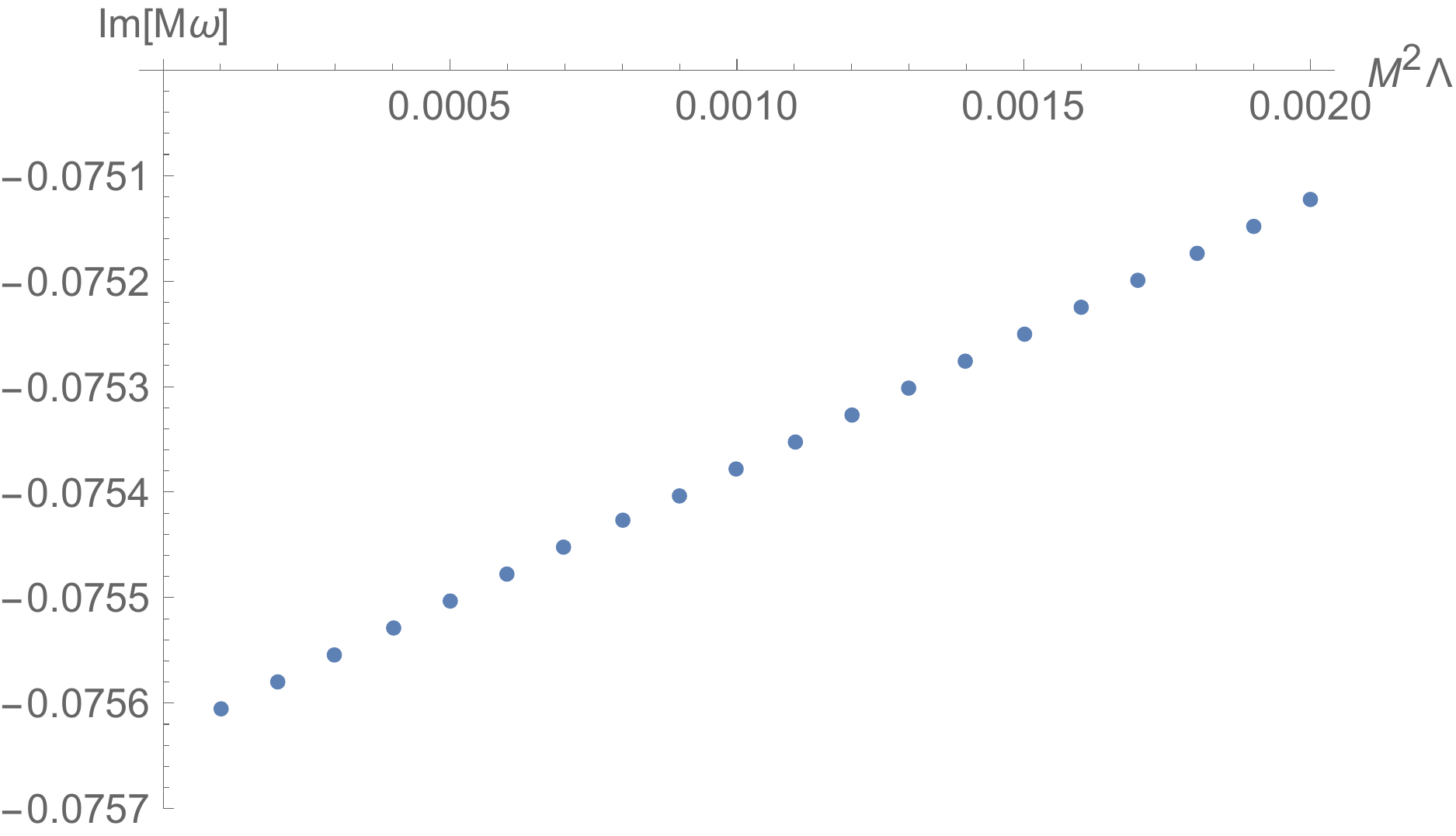}
  \end{minipage} 
\end{center}
  \caption{These panels show the real and imaginary parts of $M{}_{-2}\omega_{22}$ for almost asymptotically flat black holes at $a/M=4/5$. We can extrapolate them to the strictly flat case.}
  \label{fig:small-Lambda}
\end{figure}

A simple resolution is to extrapolate small $\Lambda$ data to $\Lambda=0$. Figure~\ref{fig:small-Lambda} shows the small $\Lambda$ behavior of $\real M \omega$ and $\im M \omega$ for $(s,\l,m)=(-2,2,2)$ and $a/M=4/5$ ranging $1/10000 \leq M^2 \Lambda \leq 20/10000$. The procedure in this note works for quite small values of $M^2 \Lambda$. We extrapolate these data to $M^2 \Lambda=0$ by using `\texttt{Fit}' with an ansatz 
\begin{equation}
\begin{aligned}
M{}_s\omega_{\l m}\approx \sum_{j=0}^{o} {}_s w_{\l m}^{(j)}(M^2\Lambda)^j.
\end{aligned}
\end{equation}
We dynamically vary the maximal order $o$, and read off stable digits of ${}_{s}w^{(0)}_{\l m}$.\footnote{Almost the same result is obtained by Richardson's extrapolation \cite{bender1978}.} We finally get
\begin{equation}
\begin{aligned}
{}_{-2}w^{(0)}_{22}(a/M=4/5)=0.58601697490887032192 - 0.07562955235606141395 i,
\end{aligned}
\end{equation}
which agrees with the QNM frequency of the asymptotically flat Kerr black holes \cite{zotero-589},
and gives a more precise prediction.

Another possible resolution has been proposed in \cite{fiziev2006, fiziev2010}. In this approach, a confluent Heun solution at $z=1$ is analytically continued to infinity that is an irregular singular point. It would be interesting to apply `\texttt{HeunC}' to asymptotically flat rotating black holes along this line.

\acknowledgments{
The author thanks Gleb Aminov, Alba Grassi, Akihiro Ishibashi, Michio Jimbo and Masashi Kimura for helpful discussions and/or kind correspondence. This work is supported by JSPS KAKENHI Grant No. JP18K03657.
}

\bibliographystyle{amsmod}
\bibliography{QNM}

\end{document}